\documentclass{raa_twocolumn}           

\usepackage{graphicx,times}             
\usepackage{natbib}
\usepackage{amssymb,amsmath}
\bibpunct{(}{)}{;}{a}{}{,}
\usepackage[pagebackref=true]{hyperref}
\usepackage{tabularx}

\begin{document}

  \title{Alert Chain and Observation Planning for Ground Wide Angle Camera Network}

   \volnopage{Vol.0 (202x) No.0, 000--000}      
   \setcounter{page}{1}          

   \author{Xu-hui Han
      \inst{1,2, *}\footnotetext{$*$Corresponding Authors, these authors contributed equally to this work.}
   \and Pin-pin Zhang
      \inst{1, *}
   \and Yu-jie Xiao
      \inst{1, *}
   \and Li-ping Xin
      \inst{1,2}
   \and Ruo-song Zhang
      \inst{1}
   \and Lei Huang
      \inst{1}
   \and Xiao-meng Lu
      \inst{1}
   \and Hong-bo Cai
      \inst{1}
   \and Yang Xu
      \inst{1}
   \and Wen-long Dong
      \inst{1}
   \and Hua-li Li
      \inst{1}
   \and Ya-tong Zheng
      \inst{1}
   \and Jian-yan Wei
      \inst{1}
   }

\institute{
    National Astronomical Observatories, Chinese Academy of Sciences, Beijing 100101,  China; {\it hxh@nao.cas.cn} {\it ppzhang@nao.cas.cn} {\it xjy@nao.cas.cn}\\
    \and
    School of Astronomy and Space Science, University of Chinese Academy of Sciences, 101408, Beijing 101408, China \\
\vs\no
   {\small Received 202x month day; accepted 202x month day}}

\abstract{The Ground Wide Angle Camera Network (GWAC-N) is a robotic telescope network. It consists of ten wide-field core telescopes (GWAC-A) and two 60cm narrow-field rapid follow-up telescopes (GWAC-F60A/B). The primary scientific goal of GWAC-N is to detect optical counterparts of gamma-ray bursts (GRBs) discovered by the \textit{SVOM} satellite. This is achieved through synchronized monitoring with the GWAC-A array. Upon receiving a GRB trigger alert, the network conducts rapid, multi-band follow-up observations using the GWAC-F60A/B telescopes. The two-stage observation process involves many telescopes, making manual control impractical. Automated operations are therefore essential. They are realized through an integrated alert processing chain and an automated observation scheduling and dispatching mechanism. To enable this, we employ the \textit{SVOM} Follow-up Observation Coordinating Service (FOCS) and GWAC-N's Automatic Observation Management (AOM) system. This paper presents the integrated alert processing workflow. It also describes the formulation of observation strategies, and the scheduling and execution of observations enabled by FOCS and AOM.
}

   \authorrunning{X. Han, P. Zhang, Y. Xiao et al.}            
   \titlerunning{Alert Chain and Observation Planning for GWAC-N}  

   \maketitle

%
%
\section{Introduction}           
\label{sect:intro}

Identifying and characterizing the optical counterparts of Gamma-Ray Bursts (GRBs) is crucial for understanding the physics of 
relativistic jets and circumburst environments. 
Optical synchrotron emission from GRBs provides essential diagnostics for 
shock physics and particle acceleration mechanisms within these jets. 
Early-time optical afterglow emission offers critical information unavailable at other wavelengths and enables precise 
astrometric localization, which is necessary for host galaxy identification and redshift determination. 
The rapid temporal evolution of optical transients—often decaying on timescales of minutes to hours—demands 
rapid, automated follow-up systems. 
Such systems are indispensable for capturing the initial brightness rise and early decay phase, which are essential 
for discriminating between theoretical afterglow models and constraining jet parameters.

In recent years, several robotic telescope networks and wide-field survey projects have developed advanced observation scheduling systems to enable automated transient follow-up. For instance, the Las Cumbres Observatory Global Telescope (LCOGT) network employs a network of globally distributed robotic telescopes with a dynamic scheduler that allocates observations \citep{Brown2013, Saunders2014}. The Zwicky Transient Facility (ZTF) uses a real-time alert system and automated follow-up observation planning \citep{Bellm2019}. The upcoming Vera C. Rubin Observatory’s Legacy Survey of Space and Time (LSST) has developed sophisticated scheduling algorithms to optimize survey efficiency \citep{Naghib2019}. In polar regions, the AST3 telescope array also implements automated observation strategies \citep{Liu2018}. Looking forward, projects like the Sitian array (also known as the GOTTA project) aim to further advance multi-telescope coordination and automated response to transients. Its pathfinder, the Mini-SiTian Array, has already demonstrated key technologies in its first two years of operation, including a dedicated master control system for automated observation management \citep{Huang2025, Wang2025, He2025}. These efforts highlight the critical role of automated scheduling in time-domain astronomy.

The Ground Wide Angle Camera Network (GWAC-N) \citep{Wei2016, Xin2026} is a robotic telescope network. 
Its primary scientific goal is to capture and perform multi-band observations of the optical counterparts of 
gamma-ray bursts (GRBs) detected by the Space-based multi-band astronomical Variable Objects Monitor (\textit{SVOM}) 
satellite \citep{Atteia2022, Cordier2026}.  
The core of GWAC-N is the Ground Wide Angle Cameras Array (GWAC-A). It comprises 10 mounts and covers approximately 3,600 
square degrees. Through optimized field planning, GWAC-A coordinates its observations with the \textit{SVOM}/Eclairs instrument 
\citep{Godet2026}. 
This strategy allows it to monitor a large portion of the Eclairs field (approximately 8,000 square degrees). 
Consequently, it significantly increases the probability of detecting prompt optical emission.
Upon receiving a GRB alert from Eclairs, the two 60cm narrow-field telescopes (GWAC-F60A/B) 
autonomously initiate follow-up observations. This ensures continuous data collection from the 
prompt emission phase into the early afterglow.

The scale of GWAC-N makes manual control impractical. Efficient automation is therefore essential and is realized 
through an integrated alert chain and an automated scheduling mechanism. Specifically, we integrate the \textit{SVOM} Follow-up 
Observation Coordinating Service (FOCS, \citep{Han2025}) with GWAC-N’s Automatic Observation Management system 
(AOM, \citep{Han2021}). This synergy allows users to design sophisticated strategies via FOCS, while AOM intelligently 
allocates telescopes and executes observations autonomously.

In the following sections, we detail the integrated workflow of the FOCS and AOM systems, including alert processing, 
the formulation of observation strategies, and the scheduling and execution of observation plans.

\section{System Architecture}

The coordinated observations between GWAC-N and \textit{SVOM} are built upon a two-stage, automated pipeline. This integrated system, centered on the FOCS 
and AOM services, handles the entire workflow from external alert ingestion to final telescope execution, enabling end-to-end automation without 
manual intervention.

The observational sequence begins with plans from the \textit{SVOM} satellite. The \textit{SVOM} Chinese/French Science Centers (CSC/FSC) and the Mission Center 
formulate the \textit{SVOM} General Program (GP) and Target of Opportunity (ToO) observation plans. The FOCS server processes the satellite's GP/ToO 
observation plans and 
generates a matching list of GWAC-A sky fields for synchronized observation with the \textit{SVOM}/Eclairs instrument. This list is ingested by the AOM 
system, where it is merged with other observational targets of GWAC-N. AOM then performs unified scheduling and dynamically assigns the fields 
to available GWAC-A telescopes for monitoring.

The system responds to transient events through two parallel, automated alert chains. The primary chain is triggered by a GRB detection from 
\textit{SVOM}/Eclairs. The resulting alert is processed by FOCS, which formulates a follow-up observation request for the narrow-field F60A/B telescopes. 
Concurrently, an internal triggering mechanism exists: if the GWAC-A array itself detects a promising optical transient during its monitoring, 
It automatically generates a follow-up request for the F60A/B telescopes to obtain confirmation and multi-band light curves \citep{Xu2020}.

All follow-up requests—whether originating from external \textit{SVOM} alerts or internal GWAC-A detections—converge at the AOM system. AOM evaluates 
their priority, integrates them into the overall observation plan, and dynamically dispatches them to the F60A/B telescopes for execution.

\begin{figure*}
  \centering
  \includegraphics[page=1,width=1\linewidth]{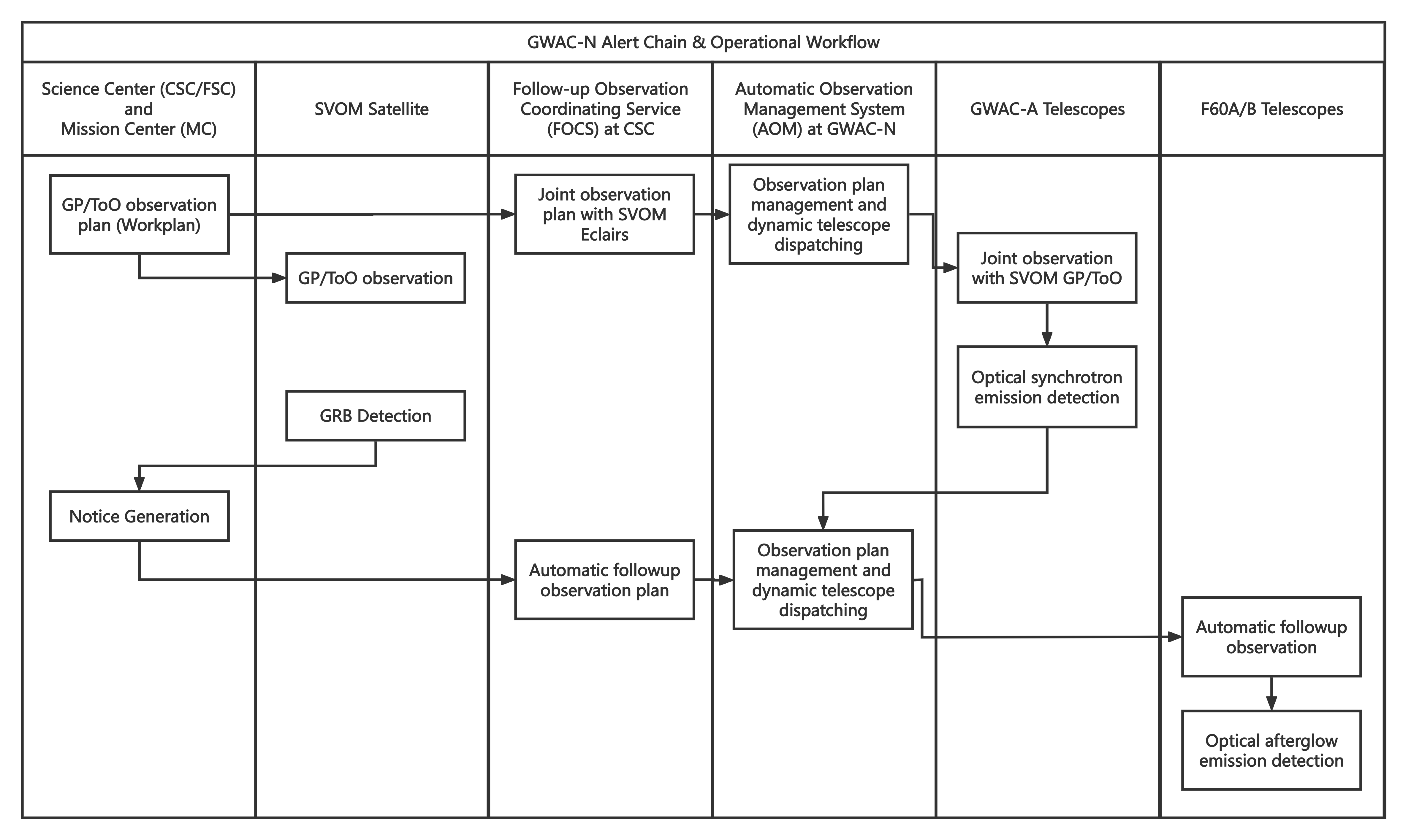}
  \caption{The end-to-end operational workflow of GWAC-N, depicting the integrated alert chain 
  and observation process. Two parallel paths are shown: (1) The processing of \textit{SVOM} General 
  Program/Target of Opportunity (GP/ToO) observation plans into joint GWAC-A observation plans; 
  and (2) The automated response chain triggered by \textit{SVOM} GRB alerts and internal GWAC-A transient 
  detections, leading to follow-up observations by the F60A/B telescopes. The FOCS and AOM systems 
  coordinate these processes, handling external alerts, planning observations, and dynamically scheduling 
  tasks across the telescope network.}
  \label{fig:workflow}
  \end{figure*}

In this architecture, FOCS and AOM play distinct, complementary roles. FOCS acts as the gateway and translator, processing all external inputs 
(GRB alerts, GP, and ToO plans) into standardized observation requests. AOM acts as the central executive and scheduler for GWAC-N. 
It oversees all observation requests, performs global observation planning, and dynamically schedules and dispatches tasks to every telescope 
in the network. The complete workflow is illustrated in Figure \ref{fig:workflow}.

\subsection{\textit{SVOM} Follow-up Observation Coordination Service}

The \textit{SVOM} FOCS is built on a Client-Server (C/S) architecture to 
coordinate multi-telescope observations. The server, hosted at the CSC, serves as the 
central hub. 
The FOCS server performs three primary functions. First, it ingests and distributes GRB alerts in real time 
through a subscription model. Second, it provides observation planning capabilities. After processing external 
inputs—such as GRB alerts, satellite's GP/ToO observation plans—it generates standardized 
observation requests. These requests are tailored to the scientific needs of users and the technical 
specifications of their instruments. Third, it runs a message notification service. This service pushes 
GRB alerts in real time to platforms like {\sc mattermost} (\footnote{\url{https://mattermost.com}}), 
{\sc enterprise wechat} (\footnote{\url{https://work.weixin.qq.com}}), 
{\sc dingtalk} (\footnote{\url{https://www.dingtalk.com}}), 
and {\sc feishu} (\footnote{\url{https://www.feishu.cn}}). This assists astronomers and 
telescope operators in rapid event validation and observation coordination.

Standardized FOCS clients enable individual telescopes or networks to connect to the server. These clients facilitate 
the exchange of alerts, requests, and plans while maintaining bidirectional links for real-time feedback. Crucially, 
the clients also possess observation planning capabilities, allowing telescope team to flexibly design customized 
strategies tailored to their specific instruments. A web-based interface provided by the client allows 
users to monitor the entire workflow in real time, from the arrival of an alert or request through to observation 
planning, execution, and feedback.

As a core ground-based system for the \textit{SVOM} mission, GWAC-N operates a customized FOCS client at its site. This client 
manages observation tasks received from the central FOCS server. Following a custom interface protocol, it generates 
specific observation requests for the GWAC-A and F60A/B telescopes and forwards them to the local AOM
system for execution. This setup supports GWAC-N's coordinated planning with \textit{SVOM} and enables its fully 
automatic GRB follow-up operations. 

The FOCS server, hosted at the \textit{SVOM} CSC, actively pushes alerts and observation plans to all registered clients, 
including the dedicated FOCS client deployed at the GWAC-N site. To ensure reliable delivery, the server maintains a queue 
for each client. In the event that a client becomes unavailable (e.g., due to network interruption or system crash), 
the server retains all pending messages. Once the client reconnects, it automatically receives the queued messages and 
sends an acknowledgment to the server, guaranteeing zero data loss during transient failures.

At the GWAC-N site, the FOCS client forwards received observation requests to the local AOM system for execution. 
The communication between the FOCS client and AOM follows a handshake protocol: after sending a request, the client 
waits for an acknowledgment from AOM. If AOM is unavailable (e.g., due to a system crash), the client does not queue 
messages for later automatic retransmission. Instead, the client's web interface displays the offline status of AOM, 
prompting human operators to manually restart the AOM system. This design avoids overloading AOM with a burst of 
pending requests upon recovery, which could otherwise cause processing delays or miss critical observation windows. 
It also prevents the execution of outdated observation plans that could pose risks to telescope safety.

\begin{figure}
  \centering
  \includegraphics[page=1,width=1\linewidth]{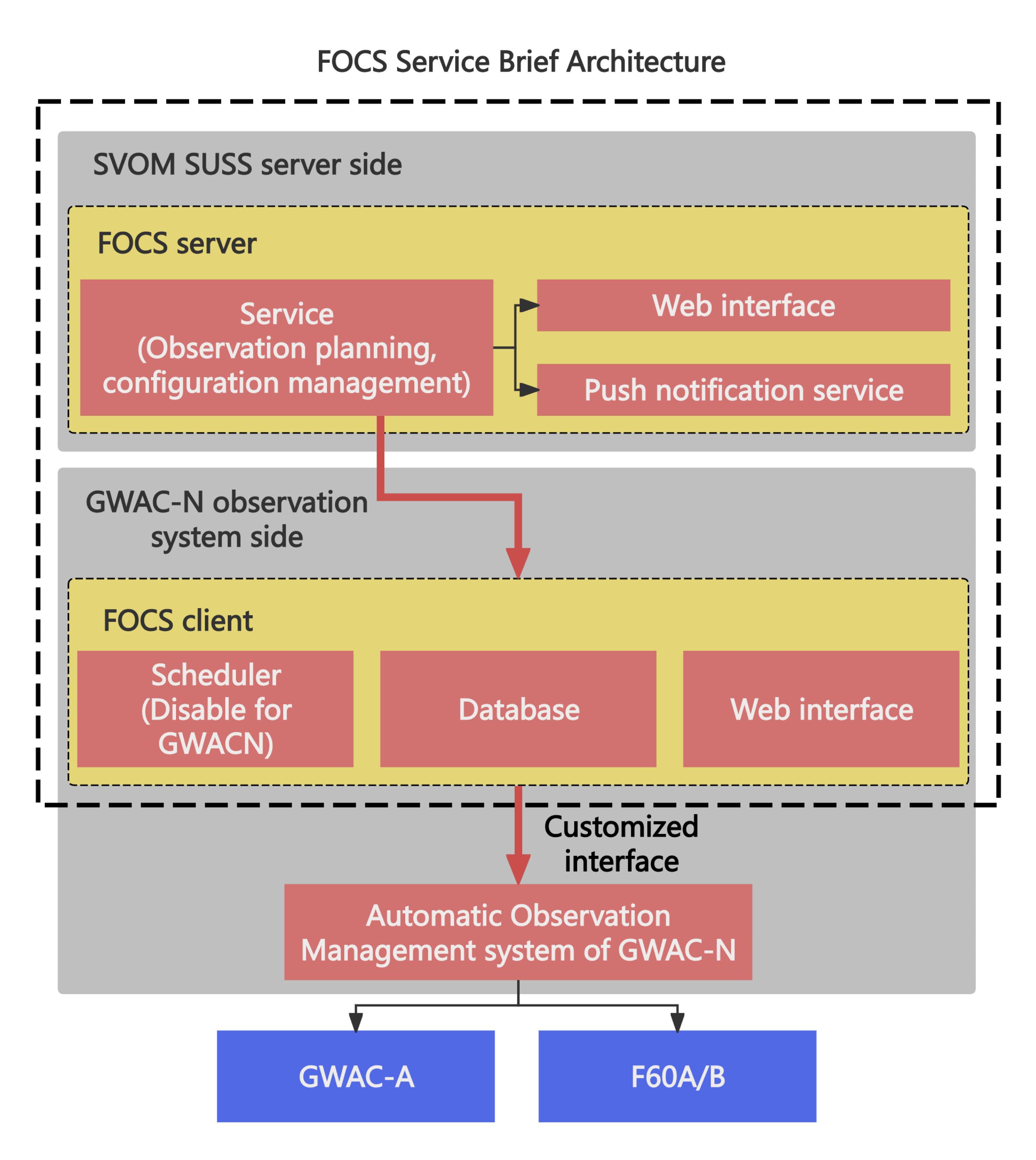}
  \caption{Architecture of the FOCS system. The server is hosted at the \textit{SVOM} CSC, 
  while a dedicated client operates at the GWAC-N site. FOCS coordinates observation planning 
  between \textit{SVOM} and GWAC-N, enabling automated follow-up observations through the GWAC-A and F60A/B 
  telescopes.}
  \label{fig:focs}
  \end{figure}

The basic architecture of the FOCS system, including its server-client relationship with GWAC-N, is 
illustrated in Figure \ref{fig:focs}.

\subsection{Automatic Observation Management System (AOM) of GWAC-N}

\begin{figure}
  \centering
  \includegraphics[page=1,width=1\linewidth]
  {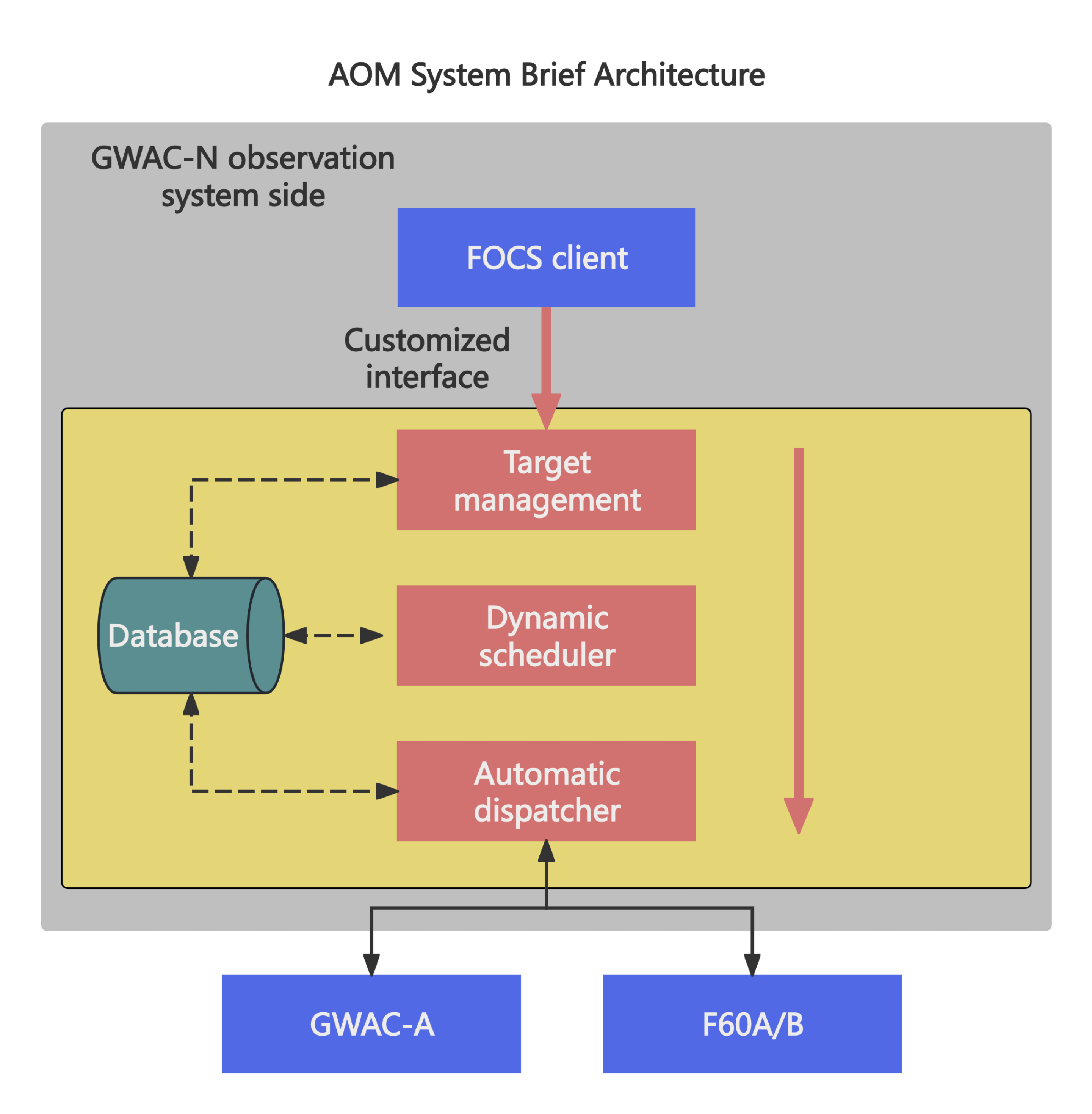}
  \caption{Schematic architecture of the Automatic Observation Management (AOM) 
  system for the GWAC-N observation system.}
  \label{fig:aom}
  \end{figure}

The architecture and workflow of the AOM system are illustrated in Figure \ref{fig:aom}. 
The AOM coordinates all observational activities within GWAC-N, managing input targets, 
distributing them to appropriate telescopes, and executing observations. Its complete architecture 
comprises several core modules: a ToO follow-up module, a target management module, 
a dynamic scheduler, an automatic dispatcher, and a communication center.

In the current implementation of AOM for GWAC-N, the ToO follow-up module—designed to process and 
standardize external GRB alerts before passing them to the target management module—is temporarily bypassed. 
This is because the FOCS system already handles the coordination, planning, and alert response for follow-up observations. 
Instead, observation requests sent from FOCS are directly ingested by the target management module in 
the form of observation targets. All targets, including those from various scientific programs of GWAC-N, 
are processed, classified, and inserted into the database by the target management module. 
The rationale for this design will be further discussed in Section \ref{discussion} of this paper.

Following target ingestion, the dynamic scheduler initiates observation planning. 
It aims to generate observation plans for all telescopes based on target prioritization and visibility. 
The highest-priority visible targets are scheduled as extensively as possible within 
the available telescope resources. Higher-priority targets may also interrupt ongoing 
lower-priority observations. Notably, GRB alerts received from FOCS are assigned the highest 
priority and can preempt routine GWAC observations. 

GRB alerts received from FOCS are assigned the highest priority and can preempt both GWAC-A and F60A/B observations. 
Preemption is implemented as an immediate abort of the current lower-priority observation, allowing the telescope 
to slew to the GRB position without delay. This ensures the earliest possible capture of the optical afterglow. 
After the high-priority observation is completed, the system may resume the interrupted observation if feasible.

The dispatcher retrieves an observation plan from the scheduler. It then checks the availability 
of the assigned telescope and verifies the target's observability. Once confirmed, it sends an 
observation command to the telescope controller and continuously monitors the observation status. 
All actions are driven by real-time information. 
A centralized communication center manages the complex interactions among external interfaces, 
internal modules, and the telescopes. It also provides message sequencing control to ensure 
that all modules and actions proceed in a predetermined order. This guarantees that the 
scheduling and dispatching operate in a fully dynamic manner.

\subsection{Interface Between FOCS and AOM}

A customized interface, developed by modifying value fields in the standard FOCS communication protocol, facilitates the 
transmission of observation requests between the FOCS client and the AOM system. Each request contains 
comprehensive metadata, including:
\begin{itemize}
    \item GRB trigger designations and trigger timestamps;
    \item Detection instruments and trigger coordinates with astrometric errors;
    \item Alert classification types, signal-to-noise ratios, and confidence levels;
    \item Designated telescope groups and observation strategies (e.g., single pointing or mosaic coverage);
    \item Observation modes (photometry or spectroscopy) and configurations (filters/gratings, slit widths, 
    exposure times, frame counts, readout speeds, observation cycles, priorities, and scheduled start/end times).
\end{itemize}
This structured dataset provides all necessary information for photometric and spectroscopic follow-up observations. 
It supports real-time candidate matching (i.e., the search and identification of optical counterparts) 
and facilitates retrospective data analysis.

\section{Observation Strategy and Planning}

\subsection{Joint Observation between GWAC-A and \textit{SVOM}/Eclairs}

The wide-field GWAC-A telescope array operates in continuous synchronization with the \textit{SVOM}/Eclairs instrument, 
monitoring the same sky region. This spatiotemporal coordination enables GWAC-A to search for prompt optical 
emission simultaneously with Eclairs’ gamma-ray detections. Real-time monitoring ensures that any optical 
counterpart appearing during the initial seconds of a GRB can be promptly identified and recorded.

Observation planning is carried out on the FOCS server. The server processes \textit{SVOM} ToO and GP observation plans—collectively 
referred to as ToO/GP workplans. For each workplan, FOCS performs several steps: it matches GWAC sky regions based on positional 
and field-of-view overlap; assigns observation priorities according to the hierarchy \textit{SVOM} ToO $>$ \textit{SVOM} GP $>$ GWAC routine 
observations; calculates observable time windows for the matched regions during the same night; and finally generates a list 
of regions as observation requests for GWAC-A.

During execution, the AOM system refines these requests by recalculating observable time windows, taking into account 
telescope pointing constraints such as hour angle, altitude limit, and lunar distance. Regions with valid observing windows 
are then sorted. The primary sorting criterion is the assigned priority. For regions with equal priority, a dedicated 
optimization algorithm determines the sequence based on a weighted score that evaluates three factors: 
proximity to the center of the Eclairs field of view (FoV), elevation, and galactic latitude. The weighting decreases 
in the order of distance, altitude, and galactic latitude, ensuring that closer regions are strongly preferred, 
while altitude and galactic latitude act as secondary tie-breakers.

The resulting sorted list is integrated into a dynamically maintained nightly observation schedule. 
This schedule is automatically re-sorted whenever the list content or the status of observations changes. 
This structured approach ensures that GWAC-A coordinates optimally with Eclairs, dynamically prioritizing 
the most scientifically critical targets while managing telescope resources and observational constraints efficiently.

The final observation sequence for equal-priority targets is determined by a composite \texttt{ranking\_score}. 
This score is a weighted sum of four observational parameters, each representing a distinct priority level designed 
to optimize field selection. The detailed formulation for each candidate GWAC field is as follows:

\noindent\textbf{Level 1: Distance Score.}
This is the primary score, based on the angular separation ($d$) between the field center and the center of the 
\textit{SVOM}/Eclairs FoV. The score increases as the distance decreases and is normalized by the matching radius $R$:
\[
S_{\text{dist}} = \text{round}\left(1 - \frac{d}{R}, 1\right), 
\\
\qquad S_{\text{dist}} \in [0, 1].
\]

\noindent\textbf{Level 2: Altitude Score and Hour Angle Score.}
The \textbf{Altitude Score} depends on the field's altitude angle ($\text{Alt}$) at the observation time. 
A higher altitude yields a higher score:
\[
S_{\text{alt}} = \text{round}\left( \frac{\text{Alt}}{90} \times 0.1, 2 \right), 
\\
\qquad S_{\text{alt}} \in [0.00, 0.09].
\]

The \textbf{Hour Angle Score} is calculated from the absolute hour angle ($|\text{HA}|$) of the field. 
It encourages observation near the meridian ($\text{HA} \approx 0$):
\[
\begin{split}
S_{\text{ha}} &= \text{round}\left( \max\left(0, (5 - |\text{HA}|) \times 0.01\right), 2 \right), 
\\
& \qquad S_{\text{ha}} \in [0.00, 0.05]
\end{split}
\]

\noindent\textbf{Level 3: Galactic Latitude Score.}
This score acts as a final micro-adjustment based on the absolute galactic latitude ($|b|$):
\[
S_{\text{gal}} = 
\begin{cases} 
0.001, & \text{if } |b| > 20^\circ \\
0.000, & \text{otherwise}
\end{cases}
\]

\noindent The final composite score is the sum of these four components:
\[
\boxed{\texttt{ranking\_score} = S_{\text{dist}} + S_{\text{alt}} + S_{\text{ha}} + S_{\text{gal}}}
\]

\textbf{Algorithm Design and Rationale}

The scoring function is hierarchically designed to enforce the strict observational priorities 
between GWAC and \textit{SVOM}/Eclairs. The weights are calibrated across distinct orders of magnitude to 
create an unambiguous ranking order:
\textbf{distance $>$ (altitude + hour angle) $>$ galactic latitude}.

The distance score ($S_{\text{dist}}$), on the order of $10^0$ (range [0, 1]), is the primary and decisive factor. 
Its dominance ensures that fields closest to the Eclairs pointing are selected first, directly maximizing 
the spatial coverage and scientific yield of the joint observations.

The altitude ($S_{\text{alt}}$) and hour angle ($S_{\text{ha}}$) scores form a secondary tier, both contributing on 
the order of $10^{-2}$. Within this tier, altitude (range [0.00, 0.09]) typically has a slightly stronger influence than 
hour angle (range [0.00, 0.05]). Together, they fine-tune the ranking by favoring targets under better observational 
conditions: higher altitude minimizes atmospheric extinction, while a smaller hour angle favors tracking near the meridian.

The galactic latitude score ($S_{\text{gal}}$) serves as a final micro-adjustment, with a value on the order of $10^{-3}$. 
It applies a subtle, systematic bias towards higher galactic latitudes ($|b| > 20^\circ$), helping to avoid crowded star 
fields and reduce background confusion.

This structured approach deliberately scales the contribution of each successive level by an order of magnitude. 
It efficiently encodes a complex multi-criteria decision into a single, sortable metric. The resulting \texttt{ranking\_score} 
enables the automated scheduler to make choices that are both scientifically targeted and observationally efficient. 
This system has been validated through the daily operation of GWAC-N, optimizing telescope time allocation to 
increase scientific throughput.

\begin{figure*}
  \centering
  \includegraphics[page=1,width=1\linewidth]
  {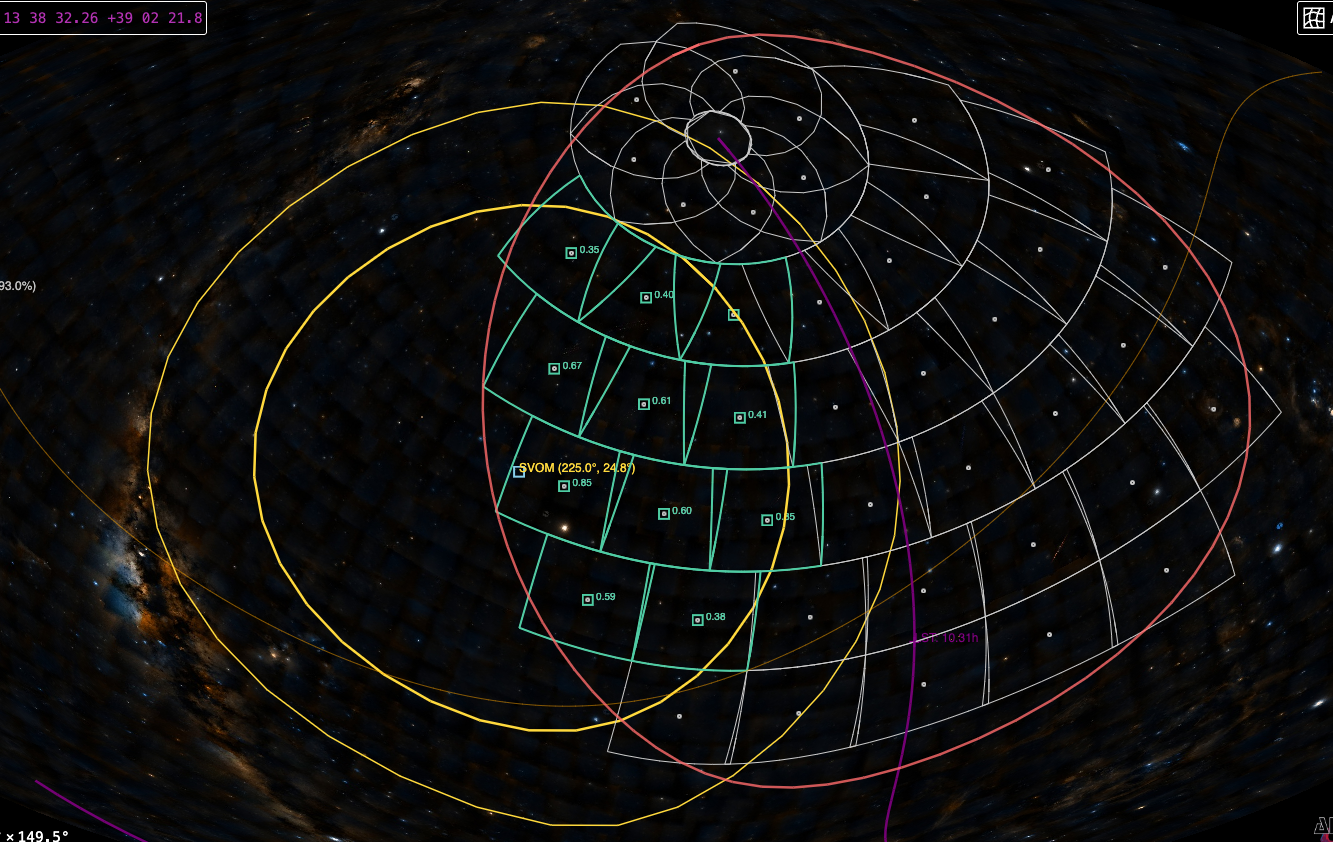}
  \caption{Web interface of the automatic co-observation planning tool. The interface displays: 
  the current SVOM pointing (small blue square) and the Eclairs FoV (thick yellow circle for central region, thin yellow circle for periphery); the GWAC-A visible sky 
  (grey box); 
  the $25^\circ$ altitude limit (red circle); the local sidereal time (purple curve); and 
  all automatically suggested fields as blue boxes, with the ten highest-scoring fields 
  labeled by their scores. The example shown corresponds to a \textit{SVOM} pointing at the source NGC~4328.}
  \label{fig:auto_tool}
\end{figure*}

The observation planning for GWAC‑N is supported by two complementary tools operating at different levels. 
An automatic co‑observation planning tool, integrated as a dedicated plugin within the FOCS server, 
implements the scoring algorithm described above. It provides observers with a real‑time web interface 
(Fig.~\ref{fig:auto_tool}) that visualizes the current \textit{SVOM} pointing, the Eclairs FoV 
(with a thick yellow circle for the central region and a thin yellow circle for the periphery, 
reflecting the uncertainty in satellite platform rotation), the GWAC‑A visible sky, and all automatically 
suggested fields as blue boxes, with the ten highest‑scoring fields labeled by their scores. 
All automatically suggested fields satisfy both temporal (visibility window overlapping the \textit{SVOM} observation) 
and spatial (overlap with the Eclairs FoV) constraints.

\begin{figure*}
  \centering
  \includegraphics[width=\linewidth]
  {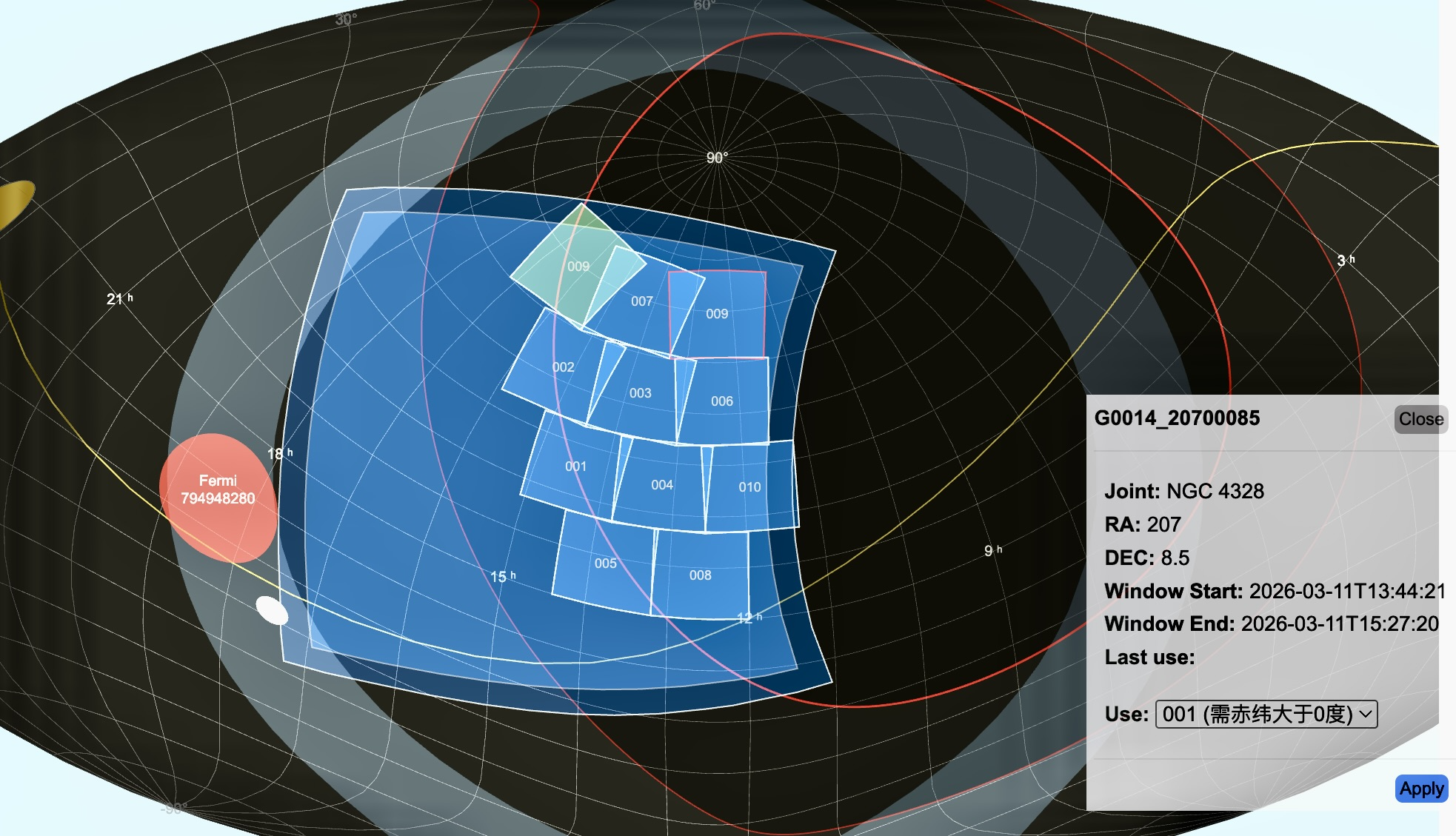}
  \caption{Interface of the manual planning tool. The main canvas shows the same sky context as Fig.~\ref{fig:auto_tool}, 
  with automatically suggested fields displayed in light green (e.g., field~009). The observer has dragged field~009 to 
  a new position (shown in red) to avoid pointing too low. The right panel provides detailed information for the selected 
  field: field number, associated \textit{SVOM} target (NGC~4328) and coordinates, visibility window, and a list of available telescopes for assignment.}
  \label{fig:manual_tool}
\end{figure*}

A manual planning tool, integrated as a plugin in the GWAC‑N observation control system, allows observers to 
adjust the field selection to address specific scientific goals or to avoid technically challenging conditions. 
Figure~\ref{fig:manual_tool} shows an example of this interface. The main canvas displays the same sky context 
as the automatic tool, including the \textit{SVOM} pointing, the Eclairs FoV, the altitude limit, and the automatically 
suggested fields (shown in white with field numbers). Observers can drag a field to a new position (the adjusted 
field appears in red) or modify its parameters via the control panel on the right. The panel provides detailed 
information for the selected field: its number, the associated \textit{SVOM} target (e.g., NGC~4328) and coordinates, 
the visibility window, and a list of available GWAC‑A telescopes that can be assigned to the observation. 
After adjustments, the updated plan is submitted to the AOM system for execution.

\begin{figure*}
  \centering
  \includegraphics[page=1,width=1\linewidth]{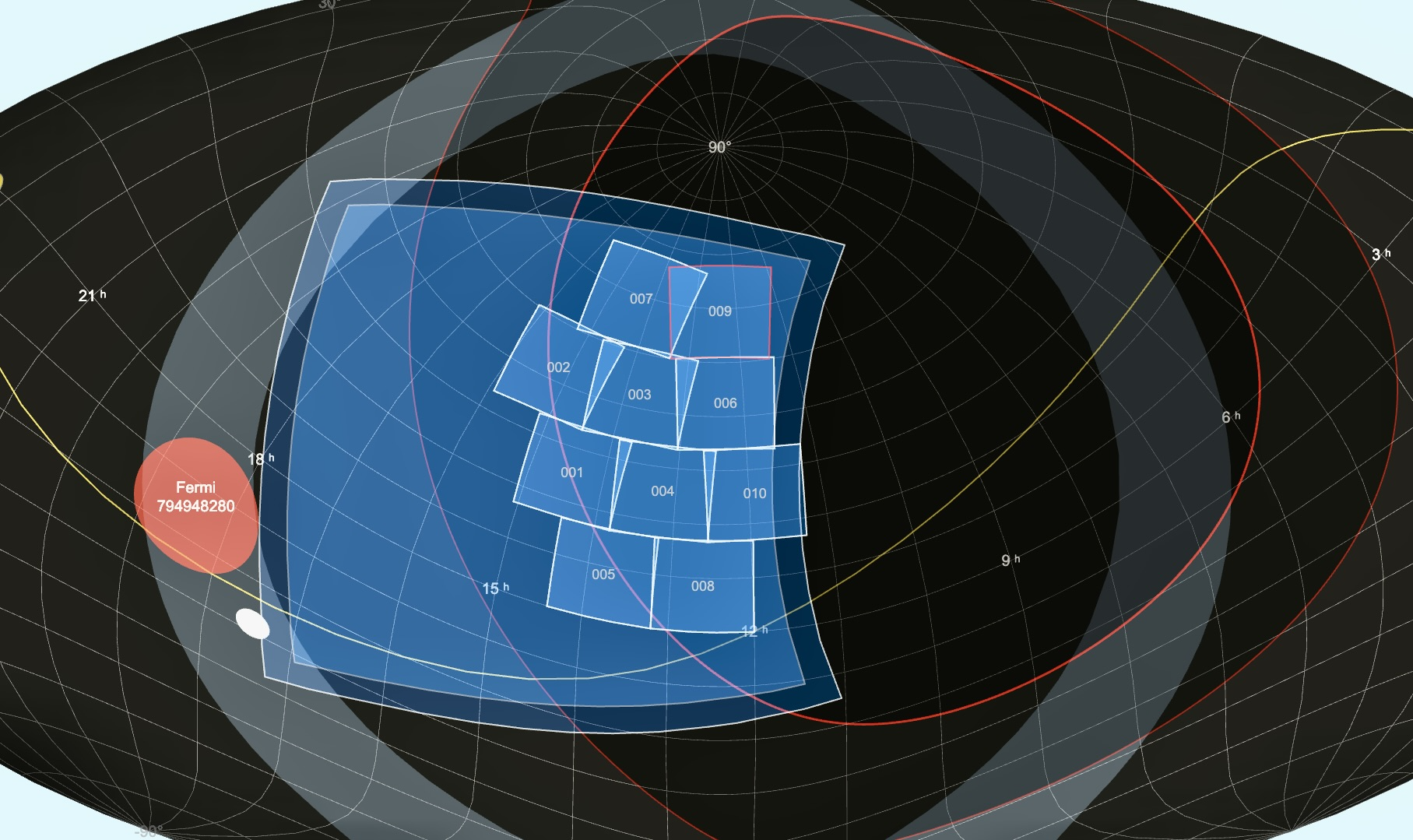}
  \caption{Visualization of the coordinated observation strategy between GWAC-A and \textit{SVOM}/Eclairs. 
  The smaller fields indicate the pointings of the GWAC-A array: white boxes represent automatically planned 
  fields, while red boxes represent manually adjusted fields (field numbers are labeled). The larger blue 
  field represents the sky coverage of the \textit{SVOM}/Eclairs instrument for source NGC~4328. The Fermi 
  alert position and its error range are also indicated. The solid red curve marks the \(25^\circ\) altitude 
  limit from the GWAC observing site. The shaded grey region highlights the area of low galactic 
  latitude (\(|b| \leq 20^\circ\)), which is strategically avoided to minimize stellar crowding and background 
  confusion. The plot demonstrates the algorithm's effectiveness in maximizing coverage of the Eclairs fields 
  within observable sky regions while adhering to observational constraints.}
  \label{fig:joint_obs}
  \end{figure*}

Figure~\ref{fig:joint_obs} presents an example of the resulting field placement for a specific \textit{SVOM} observation plan 
(source NGC~4328, RA 224.9582°, Dec 24.7562°), demonstrating the combined outcome of the automatic 
and manual planning tools. The figure illustrates how the system achieves coordinated sky coverage 
while accommodating operational considerations.

\subsection{GRB Automatic Follow-up with F60A/B}

The automated follow-up observations with the F60A/B telescopes are orchestrated by the FOCS service. 
The workflow is fully automated and includes several key steps. First, incoming GRB alerts from instruments 
such as \textit{SVOM}/Eclairs, \textit{Swift}/BAT, and \textit{EP}/WXT are filtered based on predefined criteria, primarily 
alert classification types and the signal-to-noise ratio. 
Next, visibility windows are calculated, accounting for all telescope constraints. 
Then, coordinate deviations are assessed. Finally, the event is matched to a predefined observation strategy 
to generate a detailed, executable plan.

The core of this automation is an \textbf{adaptive, time-sequenced observation strategy}. A fundamental principle is that the exposure 
sequence is not fixed but is dynamically chosen based on \textbf{the time elapsed since the GRB trigger when observations begin} 
(\(t_{\rm start}\)). This design adapts to the expected rapid decay and dimming of the optical afterglow. The strategy is divided 
into three primary scenarios, each aiming for a total on-source integration time of approximately one hour:

\begin{itemize}
    \item \textbf{Observation commencing within 3 minutes post-trigger:} Observations beginning in this prompt/early phase use an 
    increasing exposure sequence: 20s, 60s, 100s, and finally 200s. This progression prioritizes high temporal resolution 
    at the start to capture rapid early variations ({\rm e.g.} , flares or plateaus), then gradually transitions to longer exposures 
    to maintain a high signal-to-noise ratio as the source fades.

    \item \textbf{Observation commencing between 3 and 30 minutes post-trigger:} For observations starting in this early afterglow phase, 
    the strategy begins with a medium 50s exposure to achieve adequate initial depth, then switches to a series of 200s exposures. 
    This balances the need for good sensitivity with efficient monitoring of the more gradual decay.

    \item \textbf{Observation commencing beyond 30 minutes post-trigger:} During the late-time afterglow phase, 
    the source is faint and evolves slowly. The strategy employs 200s exposures exclusively to maximize detection
    depth and photometric precision for accurate light curve measurement and counterpart identification.
\end{itemize}

\begin{table*}
\centering
\setlength{\tabcolsep}{4pt} 
\caption{Adaptive Exposure Strategy for F60A/B Telescopes Based on Observation Start 
Time (\(t_{\rm start}\))}
\label{tab:f60_strategy}
\renewcommand{\arraystretch}{1.5}
\begin{tabularx}{0.95\linewidth}{|p{2.8cm}|l|l|l|X|}
\hline
\textbf{Start Time} & \textbf{Exposure Sequence} & \textbf{Frames} & \textbf{Total Time} & \textbf{Scientific Rationale} \\
\hline
\(t_{\rm start} \leq 3\) min & 20s, 60s, 100s, 200s & 20, 10, 10, 10 & 4000s & Maximize temporal resolution for early 
fast variability; adapt to rapid initial decay. \\
\hline
\(3 < t_{\rm start} \leq 30\) min & 50s, then 200s & 10, 15 & 3500s & Balance depth and efficiency for a moderately bright, 
smoothly decaying afterglow. \\
\hline
\(t_{\rm start} > 30\) min & 200s & 15 & 3000s & Maximize sensitivity and precision for a faint, slowly evolving afterglow. \\
\hline
\end{tabularx}
\end{table*}

This adaptive strategy is grounded in the typical power-law decay (\(F_{\nu} \propto t^{-\alpha}\)) of GRB afterglows. 
It offers significant advantages over a static, fixed-exposure approach. A single fixed exposure would be suboptimal: 
too long during the bright early phase, it would smear out fast variations; too short during the faint late phase, 
it would fail to detect the afterglow. The F60 strategy functions as a rule-based, dynamic sampling system. It automatically 
prioritizes high-cadence observations early on to capture dynamic behavior, then shifts to deep, sensitive exposures 
later to secure precise measurements. This ensures maximum scientific yield across the afterglow's entire 
evolutionary timeline.

\section{Performance}

We have evaluated the performance of the FOCS system between 2025 January 15 and 2025 December 25. During this period, the system 
processed 323 GRB events. It issued 2,975 observation requests to the telescope network and generated 21,936 individual 
observation tasks for coordinated observations between GWAC-A and \textit{SVOM}/Eclairs. Despite experiencing unstable network 
conditions at the Xinglong Observatory site, the system maintained an average end-to-end alert transmission delay of 
just 1.8 seconds, with a maximum delay of 5.0 seconds. It is noteworthy that the system achieved zero message loss 
throughout the entire operational period.

The observation of GRB 201223A demonstrates the operational readiness and scientific value of the integrated GWAC system \citep{Xin2022}. 
Historically, for this event, GWAC-A monitored the target field continuously with a 15-second cadence. 
Its coverage spanned from 70 minutes before the trigger to 25 minutes after, successfully capturing a fast optical 
transient synchronous with the prompt gamma-ray emission. The system now operates with an increased cadence of 3 seconds \citep{Xin2026}. 
This enhanced temporal resolution allows GWAC-A to not only study GRB afterglows in greater detail but also to 
significantly improve its sensitivity for detecting other classes of fast optical transients. Upon receiving the alert, 
the F60A telescope autonomously began follow-up observations within 45 seconds, validating the robustness of 
the automated FOCS-AOM pipeline.

Scientifically, this event underscores GWAC's unique ability to detect prompt optical emission concurrent with 
GRB high-energy activity. The resulting optical light curve shows distinct phases: an initial brightening 
($\alpha_1 \approx 0.66$), a peak at approximately 52 seconds, followed by a decay ($\alpha_2 \approx -1.07$). 
The observed optical flux was significantly in excess of the extrapolation from high-energy data. 
This characteristic, reminiscent of the physical mechanism in GRB 080319B, suggests that bright prompt 
optical emission may not be confined to the most luminous GRBs but could occur across a broader range of burst luminosities.

To quantitatively assess the effectiveness of the joint observation strategy, we plan to evaluate the detection rate 
through numerical simulations. We will develop a simulation framework that models the joint observation geometry, 
telescope visibility, and realistic operational constraints to estimate the theoretical detection rate. 
This simulated metric will provide a direct measure of the system's intrinsic capability to capture prompt optical 
emission and will serve as a quantitative benchmark for further optimization of the field selection algorithm.

\section{Discussion and Future Plans}
\label{discussion}

The GWAC-N system employs a distinct operational model within the \textit{SVOM} follow-up network. 
While observation plans for other telescopes are typically generated locally by their FOCS clients, 
planning for GWAC-N is performed centrally on the FOCS server. This server-side approach provides 
significant advantages. The server integrates comprehensive, real-time information that is challenging 
to disseminate to individual clients, including detailed \textit{SVOM} pointing plans, satellite eclipse schedules, 
South Atlantic Anomaly (SAA) entries, and the status of ongoing observational tasks. By leveraging this 
global information, the server can perform more sophisticated and optimized planning for the complex 
joint observations between GWAC-A and the \textit{SVOM} Eclairs instrument. Furthermore, this centralized 
architecture provides a robust foundation for future coordinated operations across a geographically 
distributed, multi-site GWAC-N.

A key ongoing objective is to further improve and quantitatively assess the efficiency of the joint GWAC-A and 
Eclairs observations. The primary challenge lies in maximizing the combined spatiotemporal coverage of the 
Eclairs fields within the operational constraints of the GWAC telescopes. To address this, developing well-defined, 
quantifiable efficiency metrics is crucial for refining the field-planning algorithms. Parameters such as target 
altitude, galactic latitude, and lunar distance must be intelligently weighted within the scheduling logic. 
Additionally, the timing of field transitions requires careful optimization. Together, these factors critically 
determine overall observational effectiveness and warrant systematic investigation.

Therefore, enhancing the efficiency of the GWAC-A and Eclairs joint observations remains a top development priority. 
This will be pursued by leveraging and enhancing a future multi-site GWAC-N architecture, with the goal of 
achieving a coordinated and significant increase in scientific data yield.

In addition to optimizing field selection, future work will integrate real-time weather data and observing 
conditions (e.g., cloud cover, seeing, lunar phase) into the scheduling logic. This will enable more 
robust and adaptive decision-making, improving the overall efficiency and reliability of the GWAC‑N system.

The open-source simplified client software is available for 
download\footnote{http://www.svom-gwacn.cn/doc/focs-client.zip}. To promote transparency and enable broader 
community use, we plan to release key components of the FOCS and AOM software as open-source packages in the near future.

\section{Conclusion}

This paper presents the integrated workflow and observational strategy of the GWAC-N system. 
We have demonstrated its capability to perform fully automated, two-stage coordinated observations 
in tandem with the \textit{SVOM} mission. The synergy between the FOCS and AOM systems enables efficient alert 
processing, dynamic observation planning, and rapid telescope response. Together, they form a complete, 
automated pipeline from initial GRB detection to detailed multi-band follow-up.

The system's operational readiness has been validated by observations such as GRB 201223A, which captured 
prompt optical emission and produced a well-sampled light curve revealing distinct physical phases. 
The observed optical excess suggests that bright prompt optical emission may occur across a broader range 
of GRB luminosities. This highlights GWAC-N's potential to uncover new aspects of GRB physics 
through its unique high-cadence monitoring capability.

Looking forward, the continued optimization of the field-planning algorithm remains a priority, 
and we plan to establish a quantitative benchmark by evaluating the theoretical detection rate 
through numerical simulations. Incorporating factors such as target altitude, galactic latitude, 
and lunar distance into a quantifiable efficiency metric will further enhance the spatiotemporal 
coverage of \textit{SVOM} Eclairs fields. Furthermore, the planned expansion to a geographically 
distributed, multi-site GWAC-N will significantly increase the system's sky coverage and 
observational flexibility.

With its automated, responsive architecture, the GWAC-N system stands as a powerful automated facility 
for time-domain astrophysics. It is well-positioned to contribute substantially to the study of GRBs 
and other optical transients in the multi-messenger era.

\begin{acknowledgements}
The Space-based multi-band astronomical Variable Objects Monitor (\textit{SVOM}) is a joint Chinese-French mission led by the 
Chinese National Space Administration (CNSA), the French Space Agency (CNES), and the Chinese Academy of Sciences 
(CAS). We gratefully acknowledge the unwavering support of NSSC, IAMCAS, XIOPM, NAOC, IHEP, CNES, CEA, and CNRS.
The authors are also thankful for support from 
the Strategic Priority Research Program of the Chinese Academy of Sciences (grant No. XDB0550401, XDB0550101). 
This work is also supported by 
the National Key R\&D Program of China (grant No.2024YFA1611700, 2024YFA1611702, 2023YFA1608304) . 
We thank the staffs of the GWAC-N telescope for important assistants during development and deployment of the FOCS and AOM system.
\end{acknowledgements}

\bibliographystyle{raa}
\bibliography{bibtex}

@article{Brown2013,
   author = {{Brown}, T. M. and {Baliber}, N. and {Bianco}, F. B. and {Bowman}, M. and {Burleson}, B. and {Cao}, H. and {Crellin-Quick}, A. and {Deich}, W. and {Depoy}, D. L. and {Dillon}, W. and {Dressing}, C. and {Dubberley}, M. and {Eastman}, J. D. and {Elphick}, M. and {Fish}, Z. and {Graham}, M. and {Halpern}, M. and {Handler}, J. and {Hanson}, E. and {Hidas}, M. G. and {Hovland}, R. and {Hubbard-James}, B. and {Hygelund}, J. and {Jelinek}, D. and {Jenness}, T. and {Lister}, T. A. and {Loh}, Y. S. and {McLeod}, B. A. and {Natusch}, T. and {Norbury}, M. and {Nugent}, P. and {Pickles}, A. and {Posner}, V. and {Rosing}, W. and {Riddle}, R. and {Ridenour}, M. K. and {Rodriguez}, H. and {Saunders}, E. S. and {Siverd}, R. and {Siverd}, R. J. and {Smith}, C. and {Sonneborn}, G. and {Sonneborn}, G. and {Sutherland}, W. and {Tang}, B. and {Tilleman}, T. and {Tu}, S. and {Volgenau}, N. and {Walker}, Z. and {Wold}, T.},
    title = "{Las Cumbres Observatory Global Telescope Network}",
  journal = {Publications of the Astronomical Society of the Pacific},
     year = 2013,
    month = sep,
   volume = 125,
    pages = {1031},
      doi = {10.1086/673168},
   adsurl = {https://ui.adsabs.harvard.edu/abs/2013PASP..125.1031B},
}

@article{Bellm2019,
   author = {{Bellm}, Eric C. and {Kulkarni}, S. R. and {Graham}, Matthew J. and {et al.}},
    title = "{The Zwicky Transient Facility: System Overview, Performance, and First Results}",
  journal = {Publications of the Astronomical Society of the Pacific},
     year = 2019,
   volume = 131,
    pages = {018002},
      doi = {10.1088/1538-3873/aaecbe},
   adsurl = {https://ui.adsabs.harvard.edu/abs/2019PASP..131a8002B},
}

@article{Naghib2019,
   author = {{Naghib}, Elahesadat and {Yoachim}, Peter and {Vanderbei}, Robert J. and {Connolly}, Andrew J. and {Jones}, R. Lynne},
    title = "{A Framework for Telescope Schedulers: With Applications to the Large Synoptic Survey Telescope}",
  journal = {The Astronomical Journal},
     year = 2019,
   volume = 157,
   number = 4,
    pages = {151},
      doi = {10.3847/1538-3881/aafece},
   adsurl = {https://ui.adsabs.harvard.edu/abs/2019AJ....157..151N},
}

@article{Liu2018,
   author = {{Liu}, Qiang and {Wei}, Peng and {Yu}, Ce and {Liu}, Hui-Gen and {Hu}, Yi and {Li}, Jian and {Wu}, Chao and {Lu}, Xiao-Meng and {Hao}, Jin-Xin},
    title = "{Research on scheduling of robotic transient survey for Antarctic Survey Telescopes (AST3)}",
  journal = {Research in Astronomy and Astrophysics},
     year = 2018,
   volume = 18,
   number = 1,
    pages = {5},
      doi = {10.1088/1674-4527/18/1/5},
   adsurl = {https://ui.adsabs.harvard.edu/abs/2018RAA....18....5L},
}

@article{Huang2025,
   author = {{Huang}, Yang and {Liu}, Jifeng and {Wu}, Hong and {Shang}, Zhaohui and {Luo}, Ali and {Hu}, Shaoming and {Cui}, Wenyuan and {Mao}, Yongna},
    title = "{The Mini-SiTian Array: A Pathfinder for the SiTian Project}",
  journal = {Research in Astronomy and Astrophysics},
     year = 2025,
    month = apr,
   volume = 25,
   number = 4,
      eid = {044001},
    pages = {044001},
      doi = {10.1088/1674-4527/adc795},
   adsurl = {https://ui.adsabs.harvard.edu/abs/2025RAA....25d4001H},
  eprint = {arXiv:2504.01615},
}

@article{Wang2025,
   author = {{Wang}, Zheng and {Zou}, Jin-Hang and {Ge}, Liang and {He}, Min and {Li}, Jian and {Hu}, Yi and {Tian}, Jian-Feng},
    title = "{The Mini-SiTian Array: Design and Application of Master Control System}",
  journal = {Research in Astronomy and Astrophysics},
     year = 2025,
    month = apr,
   volume = 25,
   number = 4,
      eid = {044004},
    pages = {044004},
      doi = {10.1088/1674-4527/adc78e},
   adsurl = {https://ui.adsabs.harvard.edu/abs/2025RAA....25d4004W},
}

@article{He2025,
   author = {{He}, Min and {Wu}, Hong and {Ge}, Liang and {Tian}, Jianfeng and {Wang}, Zheng and {Mu}, Haiyang and {Zhang}, Yu and {Huang}, Yang and {Zheng}, Jie and {Fan}, Zhou and {et al.}},
    title = "{The Mini-SiTian Array: First-two-year Operation}",
  journal = {Research in Astronomy and Astrophysics},
     year = 2025,
    month = apr,
   volume = 25,
   number = 4,
      eid = {044005},
    pages = {044005},
      doi = {10.1088/1674-4527/adc788},
   adsurl = {https://ui.adsabs.harvard.edu/abs/2025RAA....25d4005H},
}

@inproceedings{Saunders2014,
   author = {{Saunders}, Eric and {Lampoudi}, Sotiria and {Lister}, Tim A. and {Norbury}, Martin and {Walker}, Z.},
    title = "{Novel scheduling approaches in the era of multi-telescope networks}",
booktitle = {Observatory Operations: Strategies, Processes, and Systems V},
     year = 2014,
   series = {Society of Photo-Optical Instrumentation Engineers (SPIE) Conference Series},
   volume = 9149,
    month = aug,
      eid = {91492G},
    pages = {91492G},
      doi = {10.1117/12.2055084},
   adsurl = {https://ui.adsabs.harvard.edu/abs/2014SPIE.9149E..2GS},
}

@article{Wei2016,
   author = {{Wei}, Jianyan and {Xin}, Liping and {Zhang}, Xing and {Zheng}, Weikang and {Zhang}, Shuangnan and {Qiu}, Yulei and {Deng}, Jinsong and {Deng}, Can and {Wang}, Jing and {Wang}, Xian and {Zhou}, Xu and {Liu}, Cui and {Liu}, Zongwei and {Shang}, RenCheng and {Wang}, Chao and {Wang}, Xiang and {Wang}, Xiaofeng and {Xu}, Jialong and {Zhang}, Hui and {Zhang}, Jin and {Zhang}, Yi and {Zhang}, Zhongkai and {Zhang}, Sheng and {Zhang}, Shaohua and {Zhang}, Jianchao and {Zhang}, Jing and {Zhang}, Peng and {Zhang}, Rui and {Zhang}, Xiao and {Zhang}, Xi and {Zhang}, Xi and {Zhang}, Xia and {Zhang}, Xu and {Zhang}, Xue and {Zhang}, Yan and {Zhang}, Ying},
    title = "{The Ground Wide Angle Camera Array (GWAC): A Ground-based Monitoring System for the SVOM Mission}",
  journal = {Proceedings of the International Astronomical Union},
     year = 2016,
   volume = {12},
   number = {S325},
    pages = {275-280},
      doi = {10.1017/S1743921317000982},
   adsurl = {https://ui.adsabs.harvard.edu/abs/2016IAUS..325..275W},
}

@article{Xin2026,
   author = {{Xin}, Li-Ping and {Wei}, Jian-Yan and {Zhang}, Xing and {Zhang}, Hui and {Zhang}, Yi and {Zhang}, Zhongkai and {Zhang}, Sheng and {Zhang}, Shaohua and {Zhang}, Jianchao and {Zhang}, Jing and {Zhang}, Peng and {Zhang}, Rui and {Zhang}, Xiao and {Zhang}, Xi and {Zhang}, Xu and {Zhang}, Xue and {Zhang}, Yan and {Zhang}, Ying},
    title = "{The Ground Wide Angle Camera Network (GWAC-N): System Overview and Performance}",
  journal = {Research in Astronomy and Astrophysics},
     year = 2026,
   volume = {in press},
}

@article{Atteia2022,
   author = {{Atteia}, J. -L. and {Antier}, S. and {Choi}, C. and {Dagoneau}, N. and {Dereli}, H. and {Gogus}, E. and {Gogus}, N. and {Guidorzi}, C. and {Huang}, Y. and {Klotz}, A. and {Lachaud}, C. and {Mereghetti}, S. and {Nagataki}, S. and {Piron}, F. and {Schanne}, S. and {Takahashi}, S. and {Tanaka}, K. and {Tavani}, M. and {Wei}, J. and {Xie}, G. and {Xin}, L. and {Zhang}, B. and {Zhang}, S.},
    title = "{SVOM: A Mission to Study the Most Explosive Transients}",
  journal = {Space Science Reviews},
     year = 2022,
   volume = {218},
   number = {2},
      eid = {18},
    pages = {18},
      doi = {10.1007/s11214-022-00871-4},
   adsurl = {https://ui.adsabs.harvard.edu/abs/2022SSRv..218...18A},
}

@article{Cordier2026,
   author = {{Cordier}, B. and {Wei}, J. and {Atteia}, J. -L. and {Basa}, S. and {Claret}, A. and {Daigne}, F. and {Godet}, O. and {Goldoni}, P. and {Gogus}, E. and {Han}, X. and {Mereghetti}, S. and {Poyre}, C. and {Schanne}, S. and {Xin}, L. and {Zhang}, B.},
    title = "{The Space-based multi-band astronomical Variable Objects Monitor (SVOM) Mission}",
  journal = {Astronomy and Astrophysics},
     year = 2026,
   volume = {in press},
}

@article{Godet2026,
   author = {{Godet}, O. and {Natalucci}, L. and {Atteia}, J. -L. and {Cordier}, B. and {Schanne}, S. and {Wei}, J. and {Xin}, L. and {Zhang}, S.},
    title = "{Eclairs: The X-ray Telescope on Board the SVOM Mission}",
  journal = {Astronomy and Astrophysics},
     year = 2026,
   volume = {in press},
}

@article{Han2025,
   author = {{Han}, Xu-Hui and {Zhang}, Pin-Pin and {Xiao}, Yu-Jie and {Xin}, Li-Ping and {Zhang}, Ruo-Song and {Huang}, Lei and {Lu}, Xiao-Meng and {Cai}, Hong-Bo and {Xu}, Yang and {Dong}, Wen-Long and {Li}, Hua-Li and {Zheng}, Ya-Tong and {Wei}, Jian-Yan},
    title = "{FOCS: The Follow-up Observation Coordinating Service for the SVOM Mission}",
  journal = {Research in Astronomy and Astrophysics},
     year = 2025,
   volume = {25},
   number = {8},
      eid = {085001},
    pages = {085001},
      doi = {10.1088/1674-4527/25/8/85},
   adsurl = {https://ui.adsabs.harvard.edu/abs/2025RAA....25h5001H},
}

@article{Han2021,
   author = {{Han}, Xuhui and {Xin}, Liping and {Zhang}, Hui and {Zhang}, Yi and {Zhang}, Zhongkai and {Zhang}, Sheng and {Zhang}, Shaohua and {Zhang}, Jianchao and {Zhang}, Jing and {Zhang}, Peng and {Zhang}, Rui and {Zhang}, Xiao and {Zhang}, Xi and {Zhang}, Xu and {Zhang}, Xue and {Zhang}, Yan and {Zhang}, Ying},
    title = "{Automatic Observation Management System for the Ground-based Wide Angle Camera Array}",
  journal = {Research in Astronomy and Astrophysics},
     year = 2021,
   volume = {21},
   number = {8},
      eid = {200},
    pages = {200},
      doi = {10.1088/1674-4527/21/8/200},
   adsurl = {https://ui.adsabs.harvard.edu/abs/2021RAA....21..200H},
}

@article{Xu2020,
   author = {{Xu}, Y. and {Wei}, J. and {Xin}, L. and {Zhang}, H. and {Zhang}, Y. and {Zhang}, Z. and {Zhang}, S. and {Zhang}, S. and {Zhang}, J. and {Zhang}, J. and {Zhang}, P. and {Zhang}, R. and {Zhang}, X. and {Zhang}, X. and {Zhang}, X. and {Zhang}, X. and {Zhang}, X. and {Zhang}, Y. and {Zhang}, Y.},
    title = "{Automatic Detection of Optical Transients in GWAC Survey}",
  journal = {Research in Astronomy and Astrophysics},
     year = 2020,
   volume = {20},
   number = {11},
      eid = {176},
    pages = {176},
      doi = {10.1088/1674-4527/20/11/176},
   adsurl = {https://ui.adsabs.harvard.edu/abs/2020RAA....20..176X},
}

@article{Xin2022,
   author = {{Xin}, Li-Ping and {Zhang}, Yi and {Zhang}, Hui and {Zhang}, Zhongkai and {Zhang}, Sheng and {Zhang}, Shaohua and {Zhang}, Jianchao and {Zhang}, Jing and {Zhang}, Peng and {Zhang}, Rui and {Zhang}, Xiao and {Zhang}, Xi and {Zhang}, Xu and {Zhang}, Xue and {Zhang}, Yan and {Zhang}, Ying},
    title = "{GWAC Discovery of the Optical Afterglow of GRB 201223A}",
  journal = {The Astrophysical Journal Letters},
     year = 2022,
   volume = {925},
   number = {2},
      eid = {L18},
    pages = {L18},
      doi = {10.3847/2041-8213/ac46d1},
   adsurl = {https://ui.adsabs.harvard.edu/abs/2022ApJ...925L..18X},
}

\label{lastpage}

\end{document}